\documentclass[preprint,floats,showpacs]{elsarticle}

\usepackage{graphicx}
\usepackage{dcolumn}
\usepackage{bm}
\usepackage[english]{babel}
\usepackage[latin1]{inputenc} 
\usepackage{graphicx}
\usepackage{epstopdf}
\usepackage{amsfonts}
\usepackage{color}
\usepackage{epsfig}
\usepackage{mathrsfs}
\usepackage{amsmath,amssymb}

\renewcommand{\(}{\left(}
\renewcommand{\)}{\right)}
\renewcommand{\[}{\left[}
\renewcommand{\]}{\right]}

\begin{document}

\selectlanguage{english}

\title{Chaos and scaling in daily river flow}

\journal{Chaos, Solitons \& Fractals}

\author[rvt,els]{M. De Domenico}
 \ead{manlio.dedomenico@ct.infn.it}
 \address[rvt]{Laboratorio sui Sistemi Complessi, Scuola Superiore di Catania, Via San Nullo 5/i, 95123 Catania, Italy}
\address[els]{Dipartimento di Fisica e Astronomia, Universit\`a di Catania, Via S. Sofia 64, 95123 Catania, Italy}
  
\author[focal]{M. Ali Ghorbani}
 \address[focal]{Department of Water Engineering, Tabriz University, Tabriz, Iran}


\date{\today}

\begin{abstract}
Adequate knowledge of the nature of river flow process is crucial for proper planning and management of our water resources and environment. This study attempts to detect the salient characteristics of flow dynamics of the Karoon River in Iran. Daily discharge series observed over a period of six years (1999-2004) is analyzed to examine the chaotic and scaling characteristics of the flow dynamics. The presence of chaos is investigated through the correlation dimension and Lyapunov exponent methods, while the Hurst exponent and R\'enyi dimension analyses are performed to explore the scaling characteristics. The low correlation dimension ($2.60 \pm 0.07$) and the positive largest Lyapunov exponent ($0.014 \pm 0.001$) suggest the presence of low-dimensional chaos; they also imply that the flow dynamics are dominantly governed by three variables and can be reliably predicted up to 48 days (i.e. prediction horizon). Results from the Hurst exponent and R\'enyi dimension analyses reveal the multifractal character of the flow dynamics, with persistent and anti-persistent behaviors observed at different time scales. 
\end{abstract}



\maketitle
\begin{small}


\section{Introduction}

River flow arises as a result of interactions between climatic inputs and landscape characteristics over a wide range of spatial and temporal scales. Depending upon the nature of the climatic inputs (e.g. rainfall, temperature) and that of the landscape (e.g. basin area, slope, land use), river flow process can have regular or irregular behaviors. The degree of regularity or irregularity, i.e. \emph{complexity} (loosely speaking), of the river flow process serves as an important evaluator of its predictability. Study on the dynamic nature and predictability of river flow process has always been a key research topic in the field of hydrology and water resources and in related fields, as such play vital roles both in undertaking short-term water emergency measures and in devising long-term water management strategies.

Linearity and nonlinearity are two of the fundamental properties of river flow process that ultimately dictate its level of complexity. Although the inherent nonlinear nature of river flow process has been known for long \cite{Izzard66, Amorocho67, Amorocho71}, much of early hydrologic research (especially during 1960s-1980s) was constrained to linear approaches \cite{Harms67, Yevjevich, Klemes78, Salas81}; the lack of data and computational power largely contributed to this situation. Linear approaches continue to be prevalent in river flow and other hydrologic studies. However, advances in computational power and data collection during the last two decades or so have also facilitated proposal of nonlinear approaches as viable alternatives. The nonlinear approaches that have found widespread applications in river flow studies include artificial neural networks, data-based mechanistic models \cite{Young94}, and chaos theory \cite{Sivakumar00}, among others. The present study is concerned with chaos theory.
In the nonlinear science literature, the term \emph{chaos} is normally used to refer to situations where complex and random-looking behaviors arise from simple nonlinear deterministic systems with sensitive dependence on initial conditions \cite{Lorenz63}, and the converse also applies \cite{Goerner}. The three fundamental properties inherent in this definition, (i) nonlinear interdependence; (ii) hidden determinism and order; and (iii) sensitivity to initial conditions, are highly relevant in river systems and processes. For example: (i) components and mechanisms involved in the river system (e.g. rainfall, flow, sediment transport) act in a nonlinear manner and are also interdependent; (ii) annual cycle in river flow possesses determinism and order; and (iii) particle transport in rivers largely depend upon the time (i.e. rainy or dry season) at which the particles were released at the source, which themselves are often not known. The first property represents the 'general' nature of river flow, whereas the second and third represent its 'deterministic' and 'stochastic' natures, respectively.
In view of the obvious relevance of its fundamental concepts to river systems, chaos theory has been gaining considerable interest in river flow and related studies \cite{Sivakumar05, Koutsoyiannis06} (and Ref. therein). The outcomes of these studies are encouraging, as they reveal that chaos theory offers new avenues to study the inherent nonlinear and complex nature of the river flow process. 

The major findings in this direction are related to both chaotic and stochastic nature in the flow dynamics: it ranges from a less complex (deterministic) to a more complex (stochastic) behavior by varying the scale of aggregation. Thus, apparently controversial results may emerge by employing  analyses on river flow, because of the observed transitions from determinism to stochasticity with increasing time scale \cite{Sivakumar09}. For an extensive review on chaos theory applications to river flow (and other hydrologic) processes see Ref. \cite{Sivakumar00, Sivakumar04, Sivakumar09} (and references therein).

Hence, another important aspect in regards to the complexity of river flow process is \emph{scale}. Climatic inputs and landscape characteristics often vary in space and time scales and accordingly influence the river processes at various scales. Our general perception is that aggregation in scale averages out the variations and reduces the level of complexity. Such a perception, however, may not always stand good, since averaging may occur only within a limited range of scales, which is often defined by the processes themselves. Further, larger spatial and temporal scales may bring their own complexities, such as additional terrain types in space and climatic scenarios in time. Our knowledge and experience indicate that, for example, daily flow in a small river basin often exhibits a higher level of complexity than that of monthly flow, but the opposite is often the case when the basin size is large. The last few decades have witnessed numerous studies into the scaling properties of river networks and river flows \cite{Sivakumar04, Rinaldo06} (and Ref. therein). It is clear, from the above observations, that reliable assessment of the complexity and predictability of river flow process and identification of the appropriate models for predictions and engineering applications requires careful investigation of both the nonlinear (especially chaotic) and scaling properties. However, a close examination of the literature reveals that studies have, in general, investigated only the chaotic property or the scaling property, but not both. 

Linear analysis, by means of autocorrelation and power spectrum, and nonlinear analysis in the framework of multifractal theory have been extensively used as standard methods to investigate only the scaling feature of river flow and rainfall. It has been shown that models known as multiplicative cascades, are able to simulate river flow scaling behaviour \cite{koscielny2006long}. The range and the nature of scaling have been studied on several streamflow data from the real world, identifying basic regimes, scale breaking, universal multifractal parameters and their independence on basin size, reflecting the space-time multiscaling of both the runoff and the rainfall processes \cite{tessier1996multifractal, pandey1998multifractal}. Hence, space-time multifractals appear to be the natural framework for analyzing the scaling features of geodynamical processes including river flows, but it is worth remarking that all of these features have been investigated by using river flow and rainfall data. 
In fact, the nature of multiplicative cascades is controversial. Multifractal cascades have been successfully introduced as simple stochastic mechanisms for understanding the self-similar and intermittent behaviour of turbulent processes \cite{parisi1985turbulence, meneveau1987simple}. However, it has been shown that the deterministic variant of multiplicative cascade models can be chaotic, preserving the scaling feature \cite{biferale1994chaotic}. When, as in our case, there is no information about rainfall in the same period of river flow data, a deeper comprehension of the dynamics requires the analysis of chaotic features as well as multifractal ones: this provides the motivation for the present study to investigate both these properties in river flow.

In this paper, first, we review the main methods for investigating the chaotic behavior and the scaling properties from the time series of a process. The investigation on chaos is made by employing a multi-dimensional phase space reconstruction, using the embedding theorem \cite{Packard80, Takens81}, to obtain preliminary information on possible patterns and extent of complexity; the correlation dimension method \cite{GrassbProc83} and the largest Lyapunov exponent method \cite{Rosenstein} are used to investigate the geometry of the phase space. The scaling properties are examined through the Hurst exponent estimation \cite{Hurst51, Hurst56} and the R\'enyi dimension analysis \cite{Renyi61}. Second, as a case study, flow dynamics of the Karoon River in Iran is investigated and daily flow data, observed during 1999-2004, are analyzed by applying these methods.

\section{Methods}

In this study, the investigation of the presence/absence of chaos and scaling behaviors in the river flow series is made using a host of methods. We chose to use more than one method to avoid spurious results related to possible drawbacks of each procedure. Phase space reconstruction, correlation dimension, and Lyapunov exponent methods are employed for detecting chaos, whereas Hurst exponent and R\'enyi dimension analyses are performed to identify scaling characteristics. Brief descriptions of these methods are presented in the following.

\subsection{Phase space reconstruction}

Phase space is a useful tool for representing the evolution of a system in time. Phase space is essentially a graph or a coordinate diagram, whose coordinates represent the variables to completely describe the state of the system at any time \cite{Packard80}. The trajectories in the phase space describe the evolution of the system from some initial state, which is assumed to be known, representing the history of the system. The region of attraction of these trajectories in the phase space provides important qualitative information on the extent of complexity of the system. The idea behind such a reconstruction is that a (nonlinear) dynamic system is characterized by self-interaction, and a time series of a single variable can carry the information about the dynamics of the entire multivariable system (When multiple variables representing the system are available, it is obviously desirable to reconstruct the phase space using such multiple variables). Many methods are available for phase space reconstruction from a scalar time series. Among these, the method of delays \cite{Takens81} is the most widely used one. According to this method, for a scalar time series $\{x_{n}\}$, ($n=1,2,3,...,N$) multi-dimensional phase space can be reconstructed as
\begin{eqnarray}
\mathbf{s}_{n} = \(x_{n}, x_{n-\tau}, ..., x_{n-(m-1)\tau} \)
\end{eqnarray}
where $m$ is the dimension of the phase space, called embedding dimension, and $\tau$ is an appropriate (integer) delay time.

Many different guidelines have been offered in the literature for the selection of $m$ and $\tau$. For instance, according to the embedding theorem of Takens \cite{Takens81}, an $m = 2D_{2} + 1$-dimensional phase space is required to completely characterize a dynamic system with an attractor dimension $D_{2}$, whereas, in practice, $m > D_{2}$ would be sufficient \cite{Abarbanel90}. Similarly, for the selection of $\tau$, some studies suggest the autocorrelation function method \cite{Holzfuss}, while mutual information method \cite{Fraser86} and correlation integral method \cite{Liebert89} are also suggested. As of now, there is no general consensus on the selection of $m$ and $\tau$. In this study, the average mutual information (AMI) method is used to select $\tau$ and the false nearest neighbor (FNN) algorithm is employed to obtain $m$ following Ref. \cite{Cellucci03}, where the best values for the embedding dimension and the lag time have been identified within these approaches among the available ones.

The average mutual information (AMI) approach gets the optimal delay time $\tau$ as the first local minimum \cite{Fraser86} of the information measure
 \begin{eqnarray}
I(\tau)=\sum_{ij}p_{ij}(\tau)\log \frac{p_{ij}(\tau)}{p_{i}(\tau)p_{j}(\tau)}
\end{eqnarray}
where $p_{i}$ and $p_{j}$ are the individual probabilities of $x_{n}$ and $x_{n+\tau}$  respectively, and $p_{ij}$ is the joint probability. The AMI quantifies the amount of information about $x_{n+\tau}$ if $x_{n}$ is known: when $x_{n+\tau}$ carries the maximum information about the knowledge of $x_{n}$, AMI is locally minimum. The optimal embedding dimension $m$ can be obtained from the false nearest neighbors (FNN) search in phase space \cite{Kennel92}: the number of false neighbors in the phase space generally changes between two successive embedding dimensions $m_{0}$ and $m_{0} + 1$; the optimal embedding is realized when this number is zero. However, real time series are noisy and, hence, the dimension $m_{0} + 1$ is an optimal embedding when the percentage of false neighbors respect to the embedding $m_{0}$ is less than a certain threshold, generally fixed to be 1\%.

\subsection{Correlation dimension method}

Grassberger and Procaccia \cite{GrassbProc83} introduced a fractal measure, namely the correlation sum, to quantify the amount of correlations in a time series. Their function is defined as the number of pairs of points closer than a given distance $\epsilon$, respect to some norm $|\cdot|$, in the embedding space. An improved definition of the correlation sum, excluding $\tilde{n}$ time-correlated neighbors in the phase space, is given by
\begin{equation}
C(\epsilon,m)=\frac{2}{(N-\tilde{n})(N-\tilde{n}-1)}\sum_{i+\tilde{n}<j}\Theta\(\epsilon-|\mathbf{s}_{i}-\mathbf{s}_{j}|\)
\end{equation}
where $\Theta$ is the Heaviside step function, $\tilde{n}$ is called Theiler window and the sum is extendend to all pairs of points in the embedding space.  Correlation sum $C(\epsilon,m)$ behaves as $\epsilon^{m}$ (for any $\epsilon$) for stochastic systems and as $\epsilon^{d}$ (for values of $\epsilon$ in the scaling region) for deterministic ones and some types of coloured noise \cite{GrassbProc83, Provenzale89, Provenzale91}. When a scaling relationship between $C(\epsilon,m)$ and $\epsilon$ exists, the scaling exponent, namely, the \emph{correlation dimension} $D_{2}(\epsilon, m)$, is defined as
\begin{eqnarray}
D_{2}(\epsilon,m)=\frac{\partial \log C(\epsilon,m)}{\partial \log \epsilon}
\end{eqnarray}
For Takens embedding theorem, correlation dimension is an estimation of the degrees of freedom of the underlying process, i.e. it quantifies the number of equations needed to describe the phenomenon if it is deterministic. Numerically, $D_{2}(\epsilon,m)$ can be obtained by plotting $\ln C(\epsilon,m)$ versus $\ln\epsilon$ for different $m$ and $\epsilon$, and by taking the local slopes: for values of $m$ greater than the optimal embedding needed for phase space reconstruction, $D_{2}$ is expected to reach a plateau in the scaling region. The local slopes approach is often subjected to poor estimation of the correlation dimension: if a scaling region exists, the Takens-Theiler estimator, defined by
\begin{eqnarray}
\label{d2takens}D_{2}^{TT}(\epsilon)=\frac{C(\epsilon)}{\int_{0}^{\epsilon}\frac{C(\epsilon')}{\epsilon'}d\epsilon'}
\end{eqnarray}
can be used to improve the accuracy of estimation. Chaotic dynamical systems takes a fractional value for $D_{2}$ while deterministic, but not chaotic, dynamical systems takes an integer value. For a wide range of stochastic dynamics, $D_{2}$ goes to infinity: unfortunately the existence of a plateau and of a finite and fractional correlation dimension is not enough to distinguish chaotic time series from some stochastic ones, as coloured noise \cite{Provenzale89, Provenzale91}.

\subsection{Lyapunov exponent method}\label{Sec-Lyap}

Processes can be characterized by their sensitivity to initial conditions. If $\delta(t)$ is the distance between two nearby orbits in the phase space at some time $t$, the evolution of the \emph{sensitivity} to initial conditions $\xi(t)=\delta(t)/\delta(0)$ shows quite different behaviors for deterministic and stochastic dynamical systems \cite{Gao}:
\begin{eqnarray}
\label{lyapdef}\frac{d\xi_{\text{Det.}}(t)}{dt} &=& \lambda_{max} \xi_{\text{Det.}}(t)\\
\label{hurstdef}\frac{d\xi_{\text{Stoc.}}(t)}{dt} &=& t^{H}
\end{eqnarray} 
where $\lambda_{max}$ is called \emph{largest Lyapunov exponent} and $H$ is called \emph{Hurst exponent}. Recent unifying approaches, strictly connected to nonextensive thermodynamics \cite{Tsallis97}, are still under investigations.

Largest Lyapunov exponent is a robust measure of the mean convergence (or divergence) rate of $\xi(t)$: for processes that exhibit chaotic behavior, nearby trajectories diverge along time and $\lambda_{max}>0$, while $\lambda_{max}\leq 0$ for deterministic, but not chaotic, dynamical systems. Thus, a positive largest Lyapunov exponent is a strong signature of chaos.

A positive $\lambda_{max}$ defines a prediction horizon $t^{*}(\alpha)=N^{*}\Delta t$, i.e. the maximum number of samples $N^{*}$ that can be predicted at a sampling time $\Delta t$ within an uncertainty $\alpha$. Real time series are affected by a measurement error $\varepsilon$: it follows that $\delta(0)=\varepsilon$.  The worst uncertainty on the prediction of the series is $\alpha=\text{max}\{x_{n}\}-\text{min}\{x_{n}\}$, i.e. it equals the whole range of the series. A desirable uncertainty should be $\alpha=1.96\varepsilon$, corresponding to $95\%$ of confidence band. From eq. (\ref{lyapdef}) it follows
\begin{eqnarray}
t^{*}(\alpha) = \frac{1}{\lambda_{max}}\log\frac{\alpha}{\varepsilon}
\end{eqnarray}
and $t^{*}(1.96\varepsilon)=\lambda_{max}^{-1}\log 1.96$. The largest Lyapunov exponent can be estimated from an observed times series with different approaches \cite{Rosenstein,Sano85,Kantz94}: in the following we will adopt Rosenstein's one \cite{Rosenstein}. In the embedded space, the distance of a reference point $\mathbf{s}_{n_{0}}$ from all the other points $\mathbf{s}_{n'}$ in its neighborhood of size $\epsilon$, is calculated and averaged over the number of neighbors $|\mathcal{U}_{\mathbf{s}_{n_{0}}}|$. This procedure, iterated for each point along an orbit of $\mathcal{N}$ samples, defines the stretching factor $S(\epsilon,m,t)$ that depends on the mean convergence (or divergence) rate of nearby trajectories:
\begin{eqnarray}
\label{stretchfactor}S(\epsilon,m,t)=\frac{\Delta t}{t}\sum_{n_{0}=1}^{t/\Delta t} \log\[ \frac{1}{|\mathcal{U}_{\mathbf{s}_{n_{0}}}|} \sum_{n'} |\mathbf{s}_{n'}-\mathbf{s}_{n_{0}}| \]
\end{eqnarray}
where $t=\mathcal{N}\Delta t$. The largest Lyapunov exponent is the slope of the linear region obtained by plotting $S(\epsilon,m,t)$ versus $t$, by keeping fixed $\epsilon$ and $m$. The above procedure avoids the estimation of the tangent map, it is fast and easy to implement and it is suitable for small and noisy data sets \cite{Rosenstein}.
However, the estimation of Lyapunov exponents may produce spurious results if long time series of high quality are not available \cite{eckmann1992fundamental} or in presence of a strong stochastic contaminating signal \cite{tanaka1998analysis}.

The estimation of the prediction horizon, as previously introduced, can be verified by making use of a popular forecasting technique based on nonlinear prediction \cite{Farmer87,Sugihara90}. The procedure is as follows. In the reconstructed phase space, find all the embedding states $\mathbf{s}_{n_{0}}$ in the neighborhood $\mathcal{U}_{\epsilon}(\mathbf{s}_{N})$ of size $\epsilon$ of the current state $\mathbf{s}_{N}$. The future states $\mathbf{s}_{n_{0}+k}$, $k$ steps ahead, of all states $\mathbf{s}_{n_{0}}\in \mathcal{U}_{\epsilon}(\mathbf{s}_{N})$ are successively used for the prediction of the measurement at time $N+k$:
\begin{eqnarray}
\hat{s}_{N+k} = \frac{1}{|\mathcal{U}_{\epsilon}(\mathbf{s}_{N})|}\sum_{\mathbf{s}_{n_{0}}\in \mathcal{U}_{\epsilon}(\mathbf{s}_{N})}s_{n_{0}+k}
\end{eqnarray}
i.e. the forecasting is obtained by averaging over all closest embedding states \cite{Kantz}. In the presence of an underlying chaotic dynamics, the forecasting error is expected to exponentially increase with the forecasting time at a rate $\lambda_{max}$, corresponding to the largest Lyapunov exponent. 
The forecast error, divided by the standard deviation of the time series, is maximum when the forecast time equals the prediction horizon.

\subsection{Hurst exponent method}

Hurst exponent, previously defined in eq. (\ref{hurstdef}), has been originally introduced for investigating the diffusion features of the Nile river \cite{Hurst51, Hurst56}, and it is widely used to detect the scaling regions and to characterize the persistence of a process. Let $\{y_{n}\}$ be the partial sum time series of the series $\{x_{n}\}$, defined as 
\begin{eqnarray}
y_{n}=\sum_{i=1}^{n}\[x_{i}-\langle x\rangle\], \quad\quad \langle x \rangle = \frac{1}{N}\sum_{j=1}^{N}x_{j}\nonumber
\end{eqnarray}
If $\nu$ is an integer delay time, the Hurst exponent $H(q)$ is obtained from the structure function
\begin{eqnarray}
\label{structfunc}S_{H}(q,\nu) &=&  \frac{1}{t/\Delta t-\nu}\sum_{n_{0}=1}^{t/\Delta t-\nu} |y_{n_{0}+\nu}-y_{n_{0}}|^{q} 
\end{eqnarray}
when $S_{H}(q,\nu)\sim \nu^{qH(q)}$: the power-law is typical of a fractal process at the time scale $\nu\Delta t$. If $H(q)$ is not a constant function of $q$, the process is said to be \emph{multifractal} \cite{Gao}. Hurst exponent is a bounded measure ($0\leq H(q) \leq 1$) characterizing the persistence of a process: 
\begin{itemize}
\item $H(q) \approx 0.5$ indicates a memoryless time series, with neither short-tem nor long-term correlation between states, typical of uncorrelated stochastic processes as white noise;
\item $0 \leq H(q)< 0.5$ indicates anti-persistence: increasing trends will be followed by decreasing ones, or viceversa, and this behavior tends to be dominant for $H\to 0$;
\item $0.5<H(q)\leq 1$ indicates persistence: there is only one persistent trend typical of processes where  diffusion is faster than simple brownian motion.
\end{itemize}
For a review about Hurst analysis application to hydrological sciences we refer to \cite{Koutsoyiannis02}. However, the above method could not be robust if applied to nonstationary signals showing evident linear or seasonal trends. Instead, methods based on detrended fluctuation analysis (DFA) have been developed to remove linear (or higher order) trends, that may exists in nonstationary signals, before performing the scaling analysis. We refer to Ref. \cite{chen2002effect} for a study of the impact of nonstationary contaminating signals on the scaling features of the original data and to Ref. \cite{koscielny2006long} for a multifractal study of river flow data by means of DFA. Of course, in the case of stationary signals, Hurst analysis and DFA agree on the scaling parameters.

\subsection{Generalized $q-$th order entropy}

Given a scalar time series $\{x_{n}\}$ defined on some set $\mathcal{D}$, let us cover this set with a partition $\mathcal{P}_{\epsilon}$ of disjoint boxes of size $\epsilon$. Let $p_{\epsilon}(x)$ be the probability distribution function of the series: $p_{\epsilon}(x_{i})$ is the probability that the series takes the value $x_{i}$ for the partition $\mathcal{P}_{\epsilon}$.

A measure of the average information, needed to specify a point with accuracy $\epsilon$, is the Shannon entropy \cite{Shannon48}
\begin{eqnarray}
\mathcal{H}_{1}(\mathcal{P}_{\epsilon})=-\left\langle\log p_{\epsilon}(x)\right\rangle = -\sum_{i}p_{\epsilon}(x_{i})\log p_{\epsilon}(x_{i})\nonumber
\end{eqnarray}
When a scaling relationship exists between the amount of information and the accuracy $\epsilon$, the scaling exponent, called \emph{information dimension}, is defined as
\begin{eqnarray}
D_{1} = \lim_{\epsilon\to 0}\frac{\mathcal{H}_{1}(\mathcal{P}_{\epsilon})}{\log \frac{1}{\epsilon}}\nonumber
\end{eqnarray}

In a similar way, the concept of entropy and dimension can be respectively generalized to the $q-$order R\'enyi entropy \cite{Renyi61}
\begin{eqnarray}
\mathcal{H}_{q}(\mathcal{P}_{\epsilon})=\frac{1}{1-q}\log\sum_{i} \[p_{\epsilon}(x_{i})\]^{q}
\end{eqnarray}
and the $q-$th order R\'enyi dimension
\begin{eqnarray}
D_{q} = \lim_{\epsilon\to 0}\frac{\mathcal{H}_{q}(\mathcal{P}_{\epsilon})}{\log \frac{1}{\epsilon}}
\end{eqnarray}
where $q\in[-\infty,\infty]$ and $H_{1}$ and $D_{1}$ are obtained if $q\to 1$. The behavior of $D_{q}$ versus $q$ defines the \emph{fractal spectrum}: the underlying process of the time series $\{x_{n}\}$ is fractal when $D_{q}$ is a constant function of $q$, otherwise it is said to be multifractal. $D_{q}$ can be numerically estimated by plotting $\mathcal{H}_{q}(\mathcal{P}_{\epsilon})$ versus $\log\frac{1}{\epsilon}$ and by taking the slope of the linear region. This type of analysis requires long time series in order to avoid spurious results.


\section{Data analysis and results}

The Karoon river, with a watershed area of $58,180 \text{ km}^{2}$, is located in southwest of the I.R. of Iran, the Khuzestan province is chosen for this study. The river lies between the city of Ahwaz ($31^{\circ} 20^{\prime} \text{ N}, 48^{\circ} 41^{\prime}\text{ E}$) and the Bahmanshir river ($30^{\circ} 25^{\prime} \text{ N}, 48^{\circ} 12^{\prime} \text{ E}$), which is about 190 km long. The Karoon river is a meandering river which supplies water for the irrigation of sugarcane cultivation projects, as well as other agricultural lands. 

River flow data, observed over a period of 6 years (from January 1999 to December 2004), are considered. Fig. \ref{ts} shows the variations of daily river flow time series for a sampling time of 1 day.

\begin{figure}[!htb]
  \begin{center}
      \includegraphics[angle=0, scale=0.3]{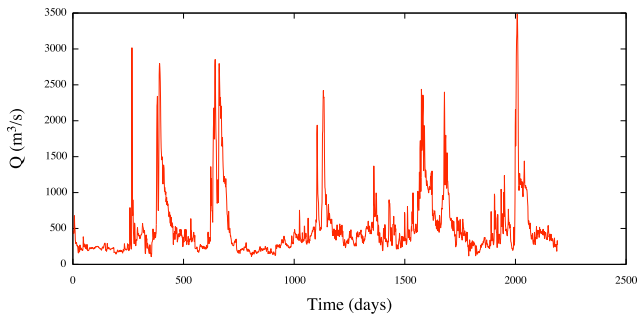}
    \caption{{Daily Karoon River flow from January 1999 to December 2004.}}
    \label{ts}
  \end{center}
\end{figure}

Optimal parameters for the phase space reconstruction, namely delay time $\tau=115$ days and embedding dimension $m=9$, are obtained from the first local minimum of AMI and FNN search, respectively. Correlation dimension $D_{2}(m,\epsilon)$ is estimated from the Takens-Theiler measure (\ref{d2takens}) up to an embedding dimension $m=18$: Fig. \ref{d2} shows a plateau for $D_{2} = 2.60 \pm 0.07$, a necessary, but no sufficient, condition for the evidence of low-dimensional chaotic dynamics, as previously discussed.  

\begin{figure}[!htb]
  \begin{center}
      \includegraphics[angle=0, scale=0.3]{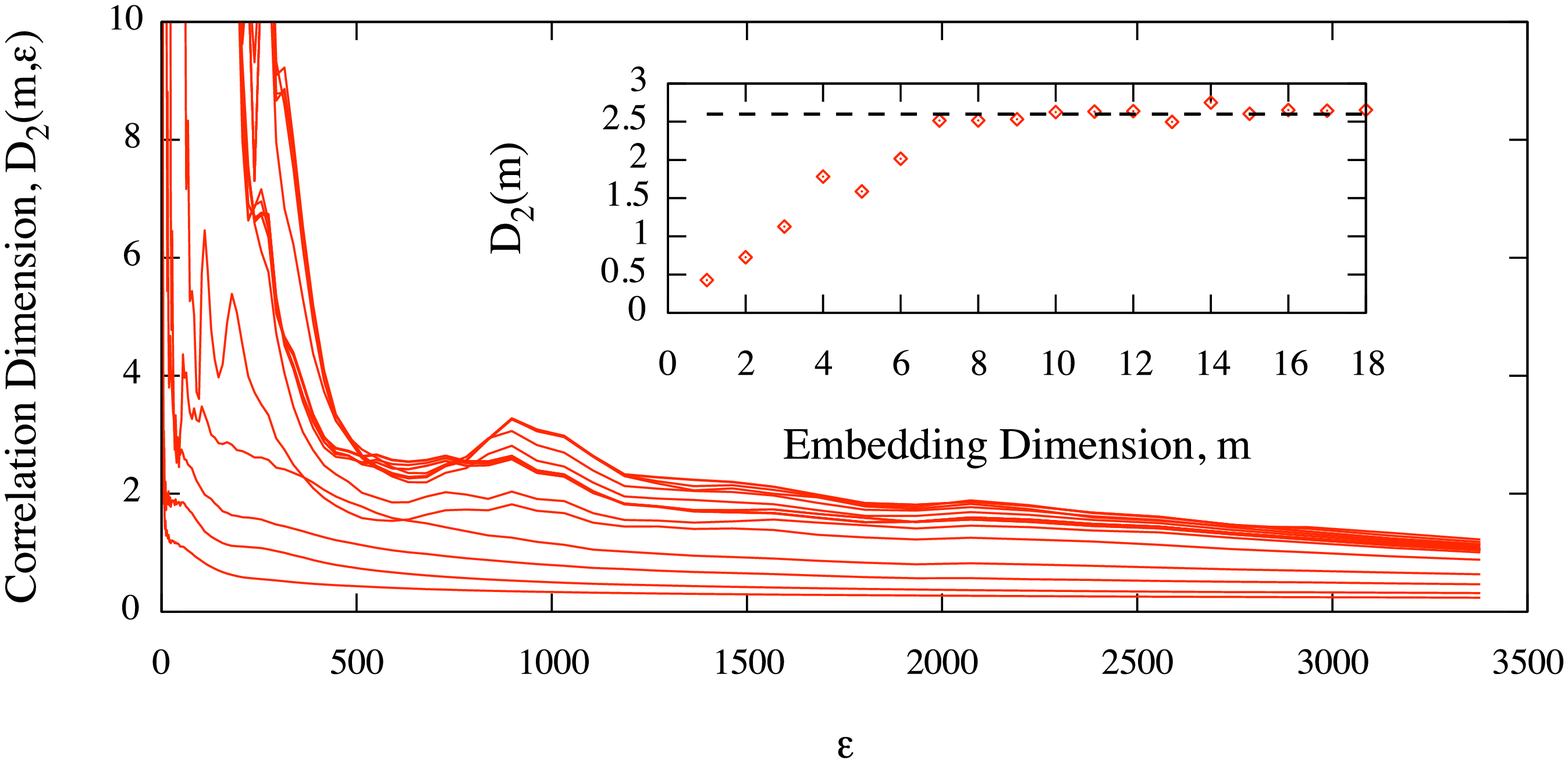}
    \caption{{Takens-Theiler estimator for the correlation dimension $D_{2}(m,\epsilon)$ and plateau fit (inner panel) corresponding to $D_{2}=2.60\pm0.07$ for $m\geq 7$ (dashed line).}}
    \label{d2}
  \end{center}
  
  \begin{center}
      \includegraphics[angle=0, scale=0.3]{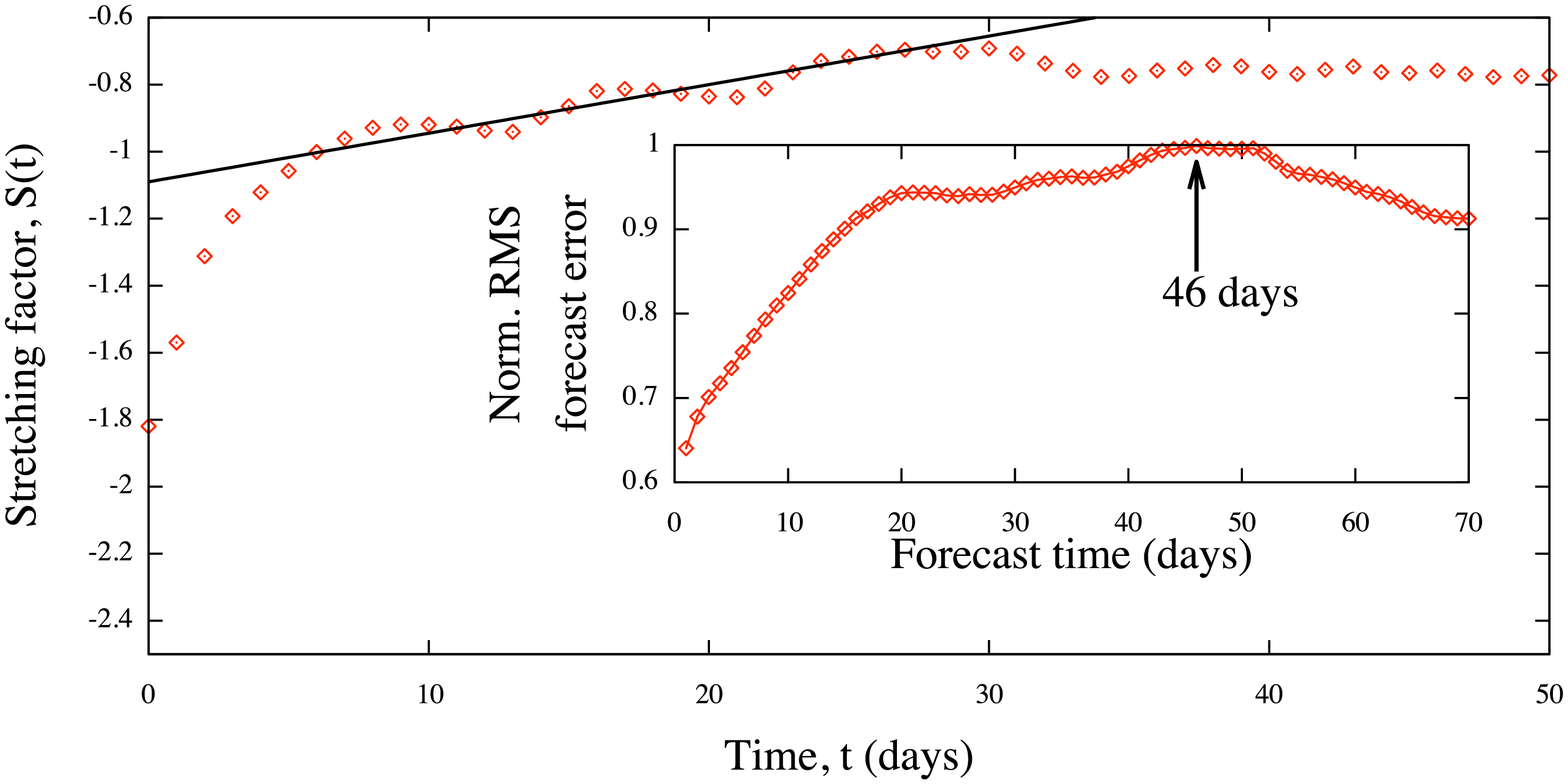}
    \caption{{Stretching factor and largest Lyapunov exponent, $\lambda_{max}=(0.014 \pm 0.001) \text{ day}^{-1}$, corresponding to the slope of the dashed line. \emph{Inner panel.} Normalized forecast error versus the forecast time: the maximum error is obtained for $t= 46$ days; the prediction horizon obtained from the Lyapunov analysis corresponds to $48.1\pm 3.4$ days.}}
    \label{lyap}
  \end{center}
\end{figure}

The stretching factor (\ref{stretchfactor}), needed for the estimation of the largest Lyapunov exponent, is shown in Fig. \ref{lyap}. We found a positive exponent, namely $\lambda_{max}=(0.014 \pm 0.001) \text{ day}^{-1}$, and a prediction horizon of $48.1\pm 3.4$ days for $\alpha=1.96\varepsilon$.

We verified the estimation of the prediction horizon by making use of the technique based on forecasting, previously described. In Fig. \ref{lyap} (inner panel) is shown the normalized forecast error versus the forecast time: the maximum error is obtained for $t= 46$ days, corresponding to a prediction horizon in excellent agreement with that one estimated from Lyapunov analysis. It is worth remarking that our estimation of both largest Lyapunov exponent and prediction horizon, characterize global features of the underlying dynamics, because they are evaluated by averages on hundreds of reference embedding states.

\begin{figure}[!htb]
  \begin{center}
      \includegraphics[angle=0, scale=0.3]{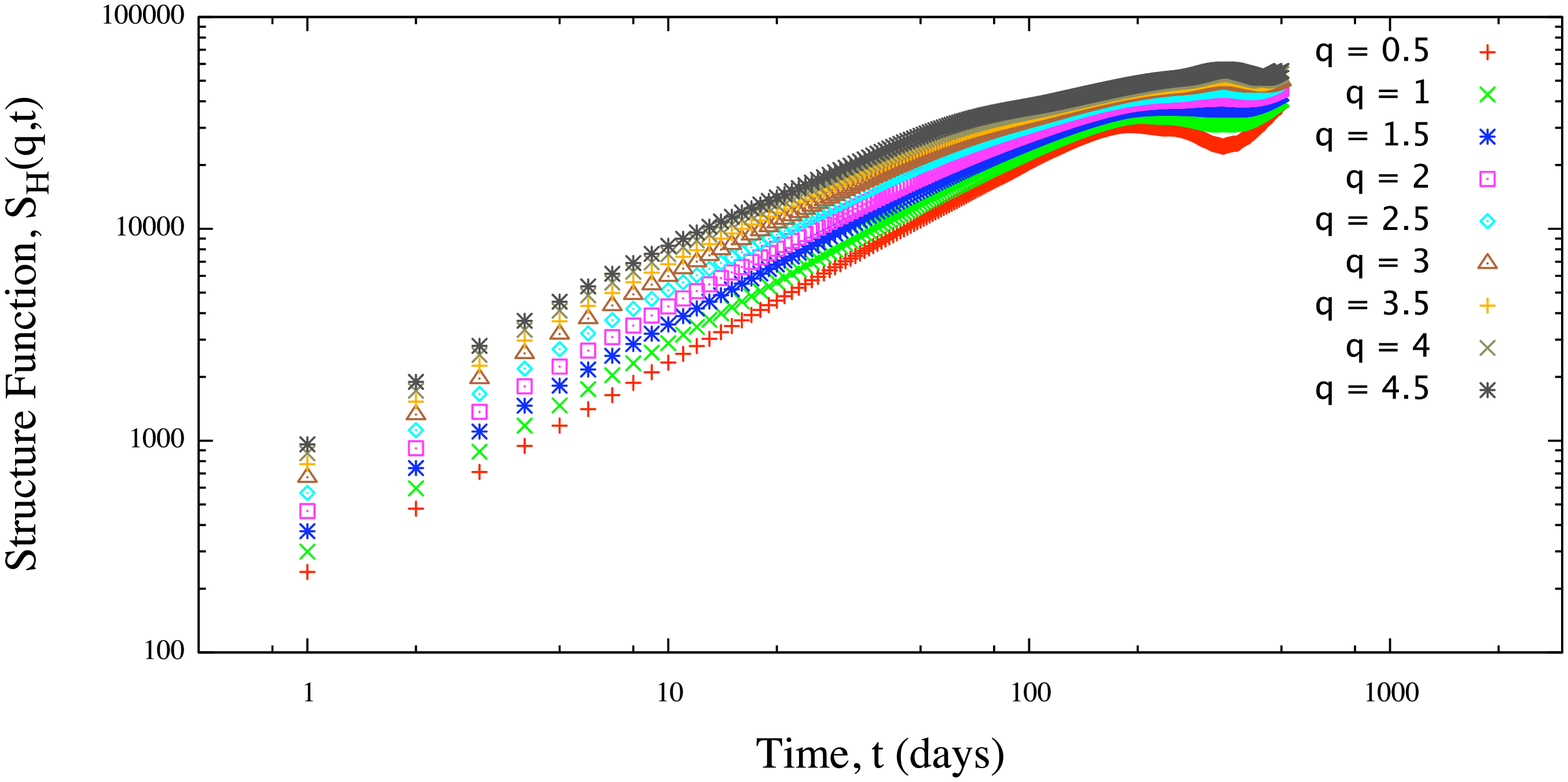}
      \includegraphics[angle=0, scale=0.3]{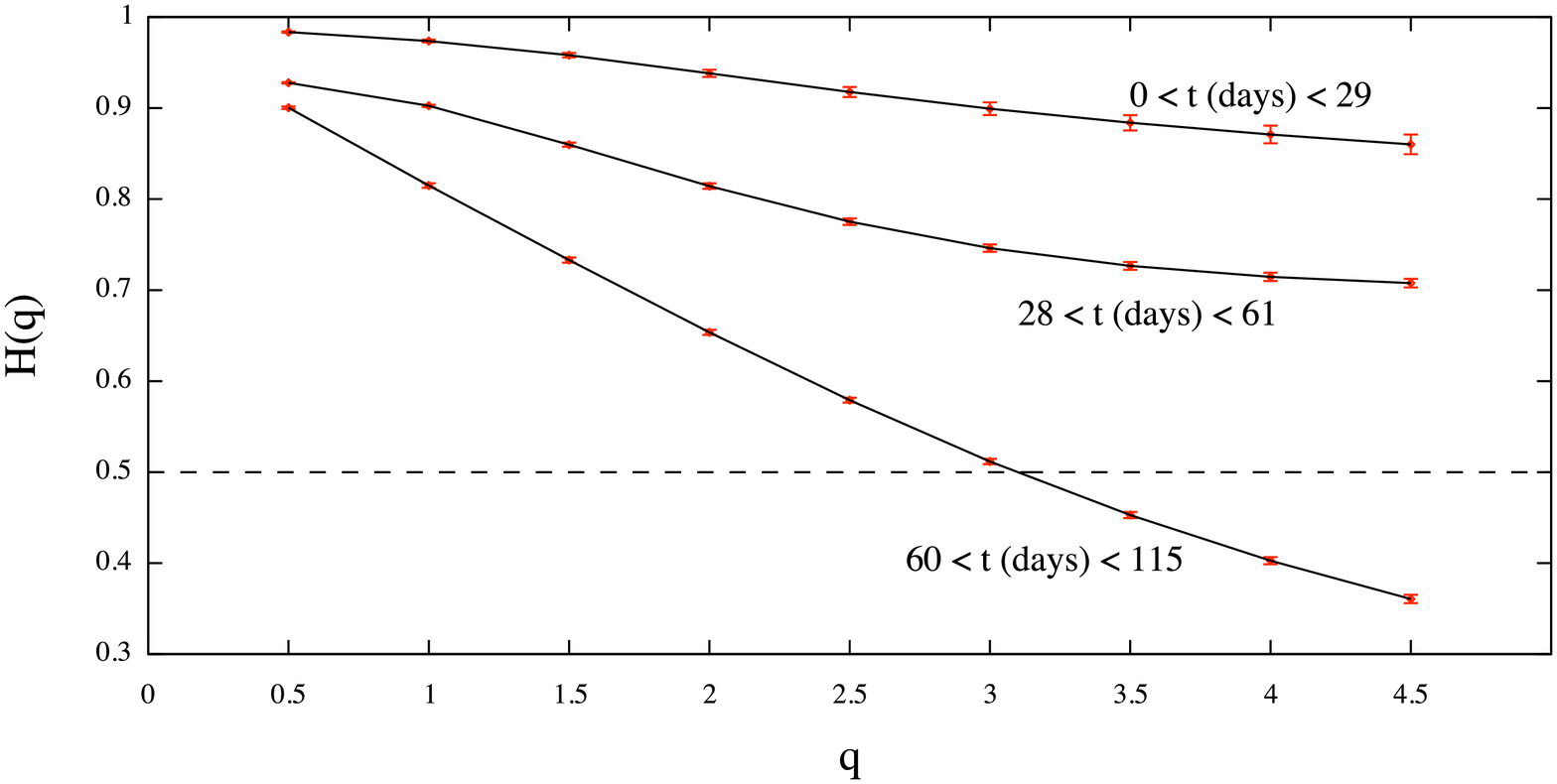}
    \caption{{Structure function $S_{H}(q,t)$ for $0.5 \leq q \leq 4.5$ (upper panel) and Hurst exponent $H(q)$ (lower panel) at different time scales. In the lower panel, the solid line corresponds to the Hurst exponent expected for uncorrelated stochastic processes.}}
    \label{hurst}
  \end{center}
\end{figure}

The structure function (\ref{structfunc}) is estimated for several values of $q$ at some different time scales, as shown in Fig. \ref{hurst} (upper panel): scaling, and thus multifractal behavior, emerges at those scales, according to a recent analysis with a multifractal detrended fluctuation method on different river flow data \cite{Movahed08}. Hurst exponents, for each $q$, are estimated from the slopes of the straight lines in Fig. \ref{hurst} (upper panel): a significant slope change defines a time scale braking. In Fig. \ref{hurst} (lower panel) Hurst exponents $H(q)$ are shown versus $q$. We identified three significant time scales from slope variations: 
\begin{enumerate}
\item From 1 to 28 days;
\item From 29 to 60 days;
\item From 61 to 114 days.
\end{enumerate}
At time scales 1) and 2), Hurst exponents depend on $q$, as for multifractal dynamical systems, $H(q)>0.5$ for all $q\in[0.5,4.5]$ and thus the time series shows a persistent behavior typical of processes exhibiting long-range dependence. At time scale 3) multifractality is stronger than previous time scales, and a transition emerges from persistent to anti-persistent behavior through a memoryless state around $q=3$, as shown in Fig. \ref{hurst} (lower panel).

Finally, the $q-$th order R\'enyi dimension $D_{q}$ is estimated for several values of $q\in[0.5,4.5]$, as shown in Fig. \ref{Renyi}: multifractal behavior emerges again, according to the Hurst exponent analysis, from an information theoretic point of view.

\begin{figure}[!t]
  \begin{center}
      \includegraphics[angle=0, scale=0.3]{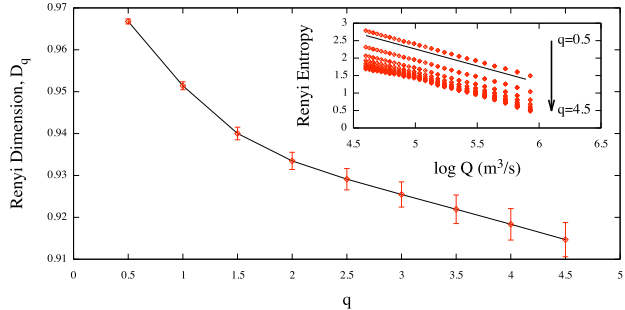}
    \caption{{The $q-$th order R\'enyi dimension $D_{q}$ estimated from generalized R\'enyi entropy (inner panel) for $0.5 \leq q \leq 4.5$.}}
    \label{Renyi}
  \end{center}  
\end{figure}

Recent studies \cite{Provenzale00, Toniolo02, Ferraris02} reveal that complex behaviors as multifractality, self-organized criticality and on-off intermittency can emerge from nonlinearly-filtered linear autoregressive processes (NFLAP). For a correct interpretations of our results, we follow the procedure suggested in Ref. \cite{Provenzale00, Toniolo02, Ferraris02}. First, we generate 1000 surrogates with the same length of the Karoon river flow time series; second, for each surrogate time series, we estimate the correlation dimension, the largest Lyapunov exponent, the R\'enyi dimensions and the Hurst exponents. Two types of surrogates, for testing two different null hypothesis, are obtained as follows:
\begin{itemize}
\item First type: a surrogate is generated by shuffling the data through a random permutation of $\{x_{n}\}$. The shuffled time series has the same probability distribution function of the original one, but time correlations are completely destroyed;
\item Second type: a surrogate is generated with the same probability distibution function and (almost) the same power spectrum of $\{x_{n}\}$. The amplitude-adjusted Fourier transform (AAFT) algorithm \cite{Theiler92,Schr-Schm96} is used to obtain this surrogate.
\end{itemize}

First type surrogates are used to test the data against the null hypothesis of an underlying uncorrelated stochastic process; second type surrogates are used to test the data against the null hypothesis of an underlying linear process distorted by a nonlinear filter.

We found no surrogates, of first or second type, showing, at the same time, a low fractional correlation dimension, a positive largest Lyapunov exponent and multifractal behavior from both R\'enyi and Hurst analyses. Thus, we can reject both the null hypotheses of an underlying uncorrelated stochastic process or an underlying linear process distorted by a nonlinear filter.


\section{Discussion}

In the last years a strong interest emerged for the underlying dynamics of river flow. Many studies reveal that both chaotic and stochastic behavior may emerge at different spatial and temporal scales. Indeed, multifractality appears to be an important feature of this process. However, each study focused or on chaotic features either on multifractal ones, but not both at the same time. In fact, deterministic and stochastic multiplicative cascades, mainly adopted as models to explain data, show similar multifractal signatures and it is not possible to deduce the real nature of the river flow without making use of both chaos and scaling analyses.

Within the present work, we investigated the salient characteristics of dynamics of the Karoon river (Iran), by examining the daily discharge time series over a period of six years (1999-2004). We followed a nonlinear approach to detect the chaotic and the scaling characteristics of the flow dynamics: the presence of chaos has been analyzed through the correlation dimension and largest Lyapunov exponent methods, while the scaling features have been explored through the Hurst and R\'enyi analyses.

Both fractional correlation dimension ($2.60 \pm 0.07$) and positive largest Lyapunov exponent  ($0.014 \pm 0.001$) suggest the presence of low-dimensional chaos: flow dynamics are dominantly governed by three degrees of freedom and can be reliably predicted up to 48 days. The estimation of the prediction horizon is in excellent agreement with the value obtained from the nonlinear forecasting analysis: the forecast error increases with the forecast time and it reaches its maximum for a time scale of 46 days. According to recent studies, our results reveal the presence of scaling typical of (chaotic) deterministic dynamical systems, although the apparently irregular behavior of the data. The fractal structure of a strange attractor emerges in the phase space and, of consequence, the underlying dynamics of the Karoon river can be successfully modelled by few deterministic equations. However, in the absence of a realistic model of the river flow, future discharge can be only monthly predicted from past and current measurements. Because of our lack of information on rainfall and atmospherical data in the same region, we can not directly quantify their effect on the prediction horizon, although our results are in good agreement with recent studies employing different forecasting techniques \cite{porporato1997nonlinear, jayawardena2000noise, lisi2001chaotic, IslamMN, sivakumar2002river}.

Results from the Hurst exponent analysis avoid a memoryless phenomenon and reveal, at different time scales, persistent or anti-persistent behavior of the river flow. Indeed, nor a unique Hurst exponent neither a unique R\'enyi dimension can be attributed to the entire process, suggesting that river discharge is characterized by anomalous scaling typical of multifractal dynamics. In particular, Hurst analysis puts in evidence three time scaling regimes. The first and the second ones, corresponding to monthly and bimonthly scales, show a persistent behavior: long-range dependence dominates the underlying dynamics and diffusion is faster than a simple brownian motion. The third scale, ranging from 60 to about 115 days, is the most extended one: for $q<3$ the process is still persistent, whereas the transition from persistent to anti-persistent behavior is evident for $q>3$. Anti-persistence is symptomatic of a diffusion process slower than a standard brownian motion: in the case of Karoon river flow, it emerges in last scaling regime depending on the way we look at the data by means of $q$. Interestingly, 115 days corresponds to the same time that minimizes the average mutual information, quantifying the maximum temporal delay, between different measurements, before both can be considered no more correlated from an information theoretic point of view. Unfortunately, because of our lack of further data, we are not able to directly relate these results to rainfall or seasonal effects.
The dependence on $q$ of the R\'enyi generalized dimensions is another important signature of multifractal behavior, although it does not produce useful information on the real nature of the underlying dynamics. Our findings from scaling analyses agree, in general, with recent results on the investigation of the multifractal nature of the river flow.

Finally, we performed the same analyses on two types of surrogate time series, to test the null hypotheses that Karoon river flow is an uncorrelated stochastic process or a linear stochastic process distorted by a nonlinear filter. We found no surrogates showing, at the same time, similar chaotic and scaling characteristics of the flow dynamics and we thus rejected both the null hypotheses.


\newpage 

\addcontentsline{toc}{section}{References} 
\begin{small}
\bibliographystyle{mprsty} 
\bibliography{draft}
\end{small}
\end{small}
\end{document}